\documentclass{aa}
\usepackage{graphicx}
\usepackage{txfonts}
\usepackage{natbib}
\begin{document}
\title{X-ray observation of ULAS J1120+0641, the most distant quasar at z=7.08}

\author{A. Moretti\inst{1}, L. Ballo\inst{1}, V. Braito\inst{1}, A. Caccianiga\inst{1}, R. Della Ceca\inst{1}, R.Gilli\inst{2}, R. Salvaterra\inst{3}, P. Severgnini\inst{1},  C. Vignali \inst{4} }  
\offprints{alberto.moretti@brera.inaf.it}
\institute{
INAF, Osservatorio Astronomico di Brera, Via Brera 28, 20121, Milano, Italy
\and
INAF, Osservatorio Astronomico di Bologna, Via Ranzani 1, 40127, Bologna, Italy
\and
INAF, Via E. Bassini 15, 20133 Milano, Italy
\and
Dipartimento di Astronomia, Universit\`a degli Studi di Bologna, Via Ranzani 1, 40127 Bologna, Italy
%
}
\date{Received ; accepted }
\date{Received ; accepted }
\titlerunning{Observation of ..}
\authorrunning{Moretti et al.}  
\abstract{} 
{We aim at probing the emission mechanism of the accreting super massive black holes in the high redshift Universe.}
{We study the X-ray spectrum of  ULAS1120+0641, the highest redshift quasar detected so far at z=7.085, which has been
deeply observed (340 ks) by XMM-Newton.}
%
{Despite the long integration time the spectral analysis is limited by the poor statistics, with 
only 150 source counts being detected.
We measured the spectrum in the 2-80 keV  rest-frame (0.3-10 keV observed) energy band. Assuming a simple 
power law model we find a 
photon index of 2.0$\pm$0.3 and a luminosity of  6.7$\pm$0.3  10$^{44}$ erg s$^{-1}$ in the 2-10 keV band,
while the intrinsic absorbing column can be only loosely constrained 
(N$_{\rm H}< $10$^{23}$ cm$^{-2}$).
 Combining our measure with published data, we calculate that the X-ray-to-optical spectral index $\alpha_{\rm OX}$ is1.8$\pm$0.1,
in agreement with the $\alpha_{\rm OX}$-UV luminosity correlation valid for lower redshift quasars.}
{We expanded to high energies the coverage of the spectral energy distribution of  ULAS1120+0641.
This is the second time that a z $>$6 quasar has been investigated through a deep X-ray observation.
In agreement with previous studies of z$\sim$6 AGN samples, we do not find
any hint of evolution in the broadband energy distribution.
Indeed from our dataset ULAS 1120+0641 is indistinguishable from the population of optically bright quasar at lower redshift.}
\maketitle
\section{Introduction}
High redshift AGNs are important probes of the Universe 
 at the end of the Dark Ages, before or around the time when the first stars formed.
Characterising their multi-wavelength properties allows us to investigate the formation and early evolution  of the super massive black holes  (SMBHs) and their  interaction with the host galaxies. 

In the last decade wide area optical-IR photometric surveys succeeded in finding a statistically significant number ($\sim$50)  
of AGNs at redshift 5.7$<$z $<$6.4 \citep{Fan06, Jiang09, Willott10a}. 
Complementing these data with IR spectroscopy for a subsample of 10 quasars,
\cite{Willott10b} derived the mass function  at z$>$6, which can be used to constrain models of SMBH evolution 
\citep{Dimatteo08, Marconi08, Shankar09, Volonteri10}. 
Recently, \cite{Venemans13} discovered three z$>$6.5 quasars in the optical/IR VIKING 332 deg$^2$ survey,
setting a lower limit  of 1.1 Gpc$^{-3}$  on the density of SMBH with mass larger than 10$^9$  M$_{\odot}$ at  6.5$<$z$<$7.5 .

\begin{figure}
\includegraphics[width=\columnwidth]{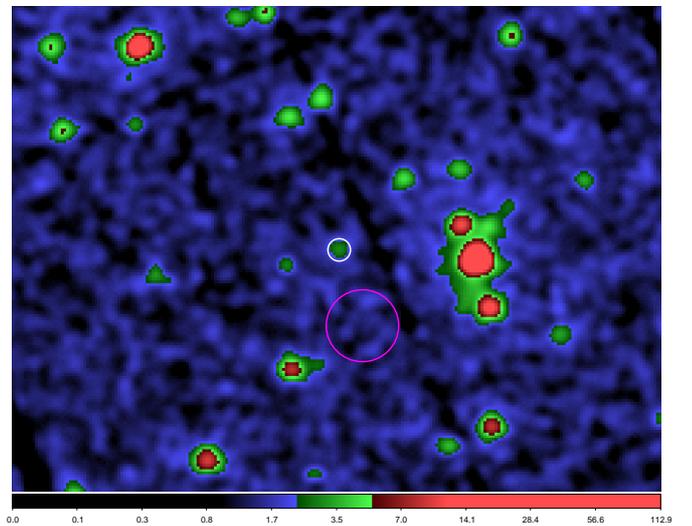}   
\caption{The 4\arcsec ~ kernel smoothed XMM-Newton MOS image of ULAS1120+641 region of the sky in the 
(observed) 0.5-2.0 keV band.  The white and magenta circles show the source and background extraction
 regions of 10\arcsec and 32\arcsec radius, respectively.}
\label{fig:ima}
\end{figure}

Few of these high redshift sources have been studied through their X-ray emission. 
Three have been observed for 10-30 ks by Chandra with 5-20 photons collected for each one \citep{Shemmer06}.
For the only SDSS J1030+0524 at z=6.3 a deep (100 ks) XMM-Newton observation has been carried out and
a spectroscopic study has been reported \citep{Farrah04}.
The statistical X-ray properties of  5$<$z$<$6 AGNs  have been  studied so far only  by stacking samples of several sources 
\citep{Vignali05,Shemmer06}.
Overall these studies indicated that the quasar broadband energy distribution has not significantly 
evolved over cosmic time, at least out to z=6.
Indeed, the mean X-ray spectral slope at z $>$5 is indistinguishable from the local population,
while the ratio between UV and X-ray  is consistent with the values observed
at lower redshift.

In this paper we discuss the deep XMM-Newton observation of  ULAS J1120+0641,
the highest redshift quasar detected so far (z=7.085). 
It has been discovered in the UK Infrared Deep Sky Survey (UKIDSS) 
and deeply studied by means of deep VLT and Gemini North spectroscopic observations \citep{Mortlock11}.
Starting from an absolute magnitude of M$_{1450,AB}$=$-$26.6$\pm$0.1 and using a correction of 4.4, \cite{Mortlock11}
estimated a bolometric luminosity of 2.4$\times$10$^{47}$erg s$^{-1}$,  while the mass of the accreting black hole
has been measured  to be 2.0$^{+1.5}_{-0.7}$ $\times$10$^9$  M$_{\odot}$ \citep{Mortlock11}. 
This implies an Eddington ratio of 1.2$^{+0.6}_{-0.5}$  \citep{Mortlock11}.
Very recently \cite{Derosa13}  refined the estimate of the black hole mass and of the bolometric luminosity by means of  
a new VLT/X-Shooter deep observation. 
They measured a mass of 2.4$^{+0.2}_{-0.2}$$\times$10$^9$ M$_{\odot}$ and an Eddington ratio of 0.5.

ULAS J1120+0641 has been also observed in the 1-2 GHz band by \cite{Momjian13} who set an upper limit
on  the ratio of the observed radio to the optical flux densities of R$<$0.5-4.3 (depending on the assumed radio spectral index alpha). Therefore, irrespective of the assumed spectral index, the source J1120+0641 at z = 7.085 is a radio quiet quasar.

\begin{table}	
\begin{tabular}{l|cccc}
\hline
Orbit  &     date     &   nom. exp. & eff. exp (MOS,PN)   \\
\hline
2281   &  May 23 2012  & 111.0   &   75.98,    69.69   \\
2294   &  Jun 18  2012  & 122.9   &   101.5,    86.95  \\
2295   &  Jun 20  2012  & 97.67   &   83.27,    64.61   \\
\hline
\end{tabular}
\caption{XMM-Newton observation log. In the third column we report the nominal exposure times in ks, while in the last column
we report the effective exposures (after data cleaning) used in the spectral analysis for both instruments.
Effective exposures are identical for MOS1 and MOS2.}
\label{tab:obs}
\end{table}
The very fact that a 2$\times$10$^9$ M$_{\odot}$ black hole exists 750 million years after the Big Bang 
strongly constrains the mass of the seed from which it developed, which cannot be lower than 5$\times$10$^5$ M$_{\odot}$ , 
unless it grew by means of thousands of merging events with massive star remnants \citep{Willott11}. 
Exploiting the detection of the [CII] emission line from the host galaxy, \cite{Venemans12} were able to estimate the star formation rate (160-440
 $_{\odot}$ yr$^{-1}$), the dust mass (0.7-6$\times$10$^8$Msun) and an upper limit on the dynamical mass of (3.6$\times$10$^{10}$(sin i)$^{-2}$ M$_{\odot}$).
While the spectrum red-ward of the Ly$\alpha$ is almost indistinguishable from lower redshift quasar, the blue-ward part
brings to an estimate of the neutral fraction of the IGM  of  $>$0.1, which is 15 times higher than z $\sim$ 6 
\citep{Mortlock11, Bolton11}.

While writing this paper, another work presenting the analysis of the same dataset, has been posted on arXiv \citep{Page13}. We will discuss the differences between their analysis and ours throughout the paper.
 
Throughout this paper  we assume H$_{\rm0}$=70 km s$^{\rm-1}$ and $\Omega_\lambda$=0.73 and $\Omega_{\rm m}$=0.27.
\begin{figure}
\includegraphics[width=\columnwidth]{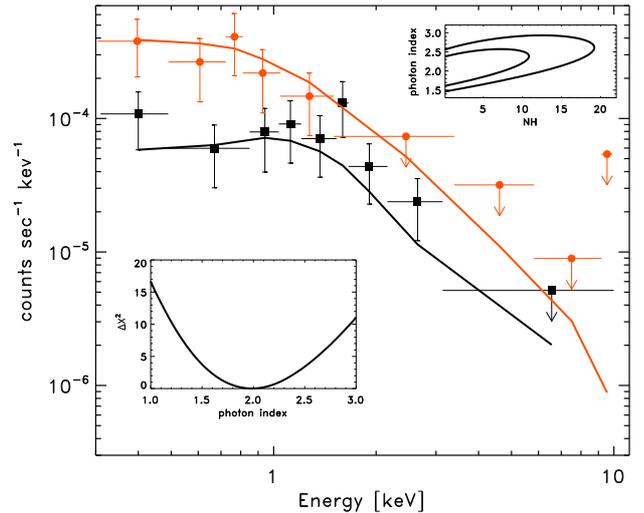}   
\caption{Black and red points and lines show the MOS and PN data and models, with the relative 1$\sigma$ errors. 
We detect significant emission up to $\sim$ 3 keV.  For the clarity of the plot, data are binned in order to have a 2$\sigma$
significance in each bin. Arrows show the 2 $\sigma$ upper limits. In the lower inset the $\Delta\chi^2$ is shown for different 
values of photon index.
In the upper inset the $\Delta\chi^2$ contour plot is shown
 as a function of  the photon index and the N$_{H}$ at the 68\% and 90\% confidence level (for 2 parameters of interest). }
\label{fig:spec}
\end{figure}
%
%

%
\section{Data} 
\label{sect:data} 
XMM-Newton observed ULAS1120+0641 in three different orbits in the period from May 24th  to June 21st 2012 for a total of  340 ks (Tab. \ref{tab:obs} ). 
The data are publicly available from the XMM science archive\footnote{\texttt{http://xmm.esac.esa.int/xsa};  PI: M. Page} .
EPIC data have been processed and cleaned using the Science Analysis Software (SAS ver 13.0.1) and analyzed
using standard software packages (FTOOLS ver. 6.13).   
The data were filtered for high background time intervals; for each observation  and EPIC camera, we  extracted the 10-12 keV light curves  and  filtered out the time intervals where the light curve was 2$\sigma$ above the mean.  For the scientific analysis we considered only events corresponding to patterns 0-12  and pattern 0-4 for the EPIC-MOS1/MOS2 and EPIC-pn, respectively.
We ended up with 260.75 and 221.25 useful ks for the MOS and the PN, respectively (Tab. \ref{tab:obs}). 

We restricted our spectral analysis to a circular region centred on the optical position of the source
 (RA: 11:20:01.48 DEC:06:41:24.3)  
with a radius of 10\arcsec, corresponding to 55\% of the encircled energy fraction at 1.5 keV
\footnote{http://xmm.esac.esa.int/external/xmm\_user\_support/documentation}.
The background was extracted from an adjacent circular region, $\sim$10 times larger (Fig.\ref{fig:ima}).
In the source region and  in the 0.3-2.0 (0.3-10) keV observed energy band, the source is 45\%  and 30\% 
(28\% and 10\%) of the total (source+background) signal, for the MOS and  the PN respectively, while using a wider extraction region would result in an unacceptable signal/background ratio.  
We extracted the data from the same regions for both the MOS detectors and for the PN, summing together the six MOS 
and the three PN spectra respectively (together with the calibration matrices).
In total the estimated source counts registered by the MOS and PN in the 0.3-10 keV band are $\sim$ 86 and 71,
95\% of which are below 3.0 and 2.0 keV respectively, (Fig.\ref{fig:spec}).

Using \texttt{Xspec} version 12.7.1 \citep{Arnaud96}, we modelled the MOS and PN spectra with a simple absorbed
power law, the absorption factor being frozen to the Galactic value (5.1$\times$10$^{20}$ cm$^{-2}$) as measured by 
the HI Galaxy map \citep{Kalberla05}. 
Spectra were grouped ensuring a minimum of one count for each bin and the best fit was calculated using the C-statistics. 
We found that the best photon index is 1.98$^{+0.26}_{-0.26}$ (errors are quoted at 68\% confidence level) and the flux 
in the 0.5-2.0 keV band is 9.3$^{+1.3}_{-1.2}$ $\times$10$^{-16}$erg s$^{-1}$ cm $^{-2}$ (Tab.\ref{tab:best}). 

Folding the model with an absorber at the same redshift of the source (\texttt{zwabs} in XSPEC), the 1 $\sigma$ upper limit 
on the intrinsic absorbing column density is 10$^{23}$ cm$^{-2}$.
Although the red-shifted iron K$_{\alpha}$ falls in a very favourable energy band (0.8 keV) both for detector energy resolution
and response, we can set only a very loose 1 $\sigma$ upper limit of 0.92 keV (rest-frame) to its equivalent width.

Using the bolometric luminosity reported by  \cite{Mortlock11}, 2.4$\times$10$^{47}$erg s$^{-1}$,
yields L$_{\rm bol}$/L$_{\rm [2-10]keV}$ of 361$^{+19}_{-15}$, where the errors account only for the 
luminosity uncertainties.

\begin{table}	
\begin{tabular}{l|lll}
\hline
   &      & &  \\
\hline
N$_{\rm H}$   & 5.1$\times$10$^{20}$    &  (frozen)          &  cm $^{-2}$                                 \\
norm.             & 4.16$\times$10$^{-7}$     &  (3.65 - 4.70)   & keV$^{-1}$s$^{-1}$cm$^{-2}$      \\ 
ph.ind.           & 1.98                                  &  (1.72 - 2.26)   &                                                  \\
\hline
Flux    [0.5-2]  keV   &  9.3  10$^{-16}$   &  (8.2 - 10.6)    &  erg s$^{-1}$ cm $^{-2}$ \\ 
Lumin [2-10]keV      &  6.7  10$^{44}$    &  (6.4 - 6.9)    & erg s$^{-1}$  \\
\hline
d.o.f.     &  637       & &  \\
C-stat    &  693      & &  \\
\hline
\end{tabular}
\caption{Best fit parameters. The uncertainty intervals reported in the second  column are at the 68\% confidence level.}
\label{tab:best}
\end{table}

 Our spectral slope best fit is lower than the one reported by \cite{Page13}, who find
2.64$^{+0.37}_{-0.33}$ (1 $\sigma$ errors). 
Although the two measures differ by less than 2 $\sigma$, 
the discrepancy is more significant considering that  the measure are not independent since we used  the same dataset.
A possible origin of this discrepancy is  a different  estimate of the background.
It is well known that the X-ray background cosmic variance produces an extra systematic term to be summed to the 
poissonian/statistical  uncertainty \citep{Moretti09}. 
Moreover the instrument component, which is a substantial fraction of the EPIC background is not 
spatially uniform  on the detector \citep{Kuntz08}. 
To check our measure against its non-poissonian variations, 
we repeated our analysis varying five times the size and the position of the background extraction region:
we  found that the five photon indexes best fit values are in the range 1.6-2.3 with a mean of 1.9.
Therefore we cannot exclude that a different background estimate is the cause of the difference 
between our spectroscopic measure with respect to the one reported by \cite{Page13}.

We note that our flux measure in the 0.5-2.0 keV band is 1.5 times ($\sim$ 2 $\sigma$) 
higher than the one reported by \cite{Page13}.
As the background in the soft band is almost two times the source, this discrepancy can be explained 
by a 20-30\% difference in the background normalization estimate. 
This, again, suggests that  the probability that the difference between the 2 measures is  due to a  background
fluctuation is not negligible.

ULAS1120+0641 has been observed by Chandra for 15.8 ks, in February 2011 (Observation-id 13203). 
We analysed the  processed (level 2) event file available from the Chandra science archive\footnote{http://cda.harvard.edu/chaser/}.
Due to the short exposure only 4 source counts have been collected in the 0.5 -2.0 keV band.
Assuming the spectral parameters measured by  XMM-Newton,  we would expect 2.8 counts in a 15.8 ks Chandra observation,
consistent with the observed data.
Thus, at variance with  \cite{Page13},  we do not find a significant variation in the flux level of 
ULAS1120+0641, between the Chandra and XMM observations.

\section{Discussion}
\subsection{X-ray photon index} 
\label{sect:xgam} 

According to the AGN standard model, X-ray photons are produced by the inverse Compton scattering 
of the accretion disk UV photons by the electrons in the corona \citep{Haardt91}.
Therefore the X-ray spectrum can be used as a direct probe of the physical mechanisms acting very 
close to the black hole.
In particular, the slope of the X-ray spectrum gives direct information about the energy distribution of the electrons
in the corona.
\begin{figure}
\includegraphics[width=\columnwidth]{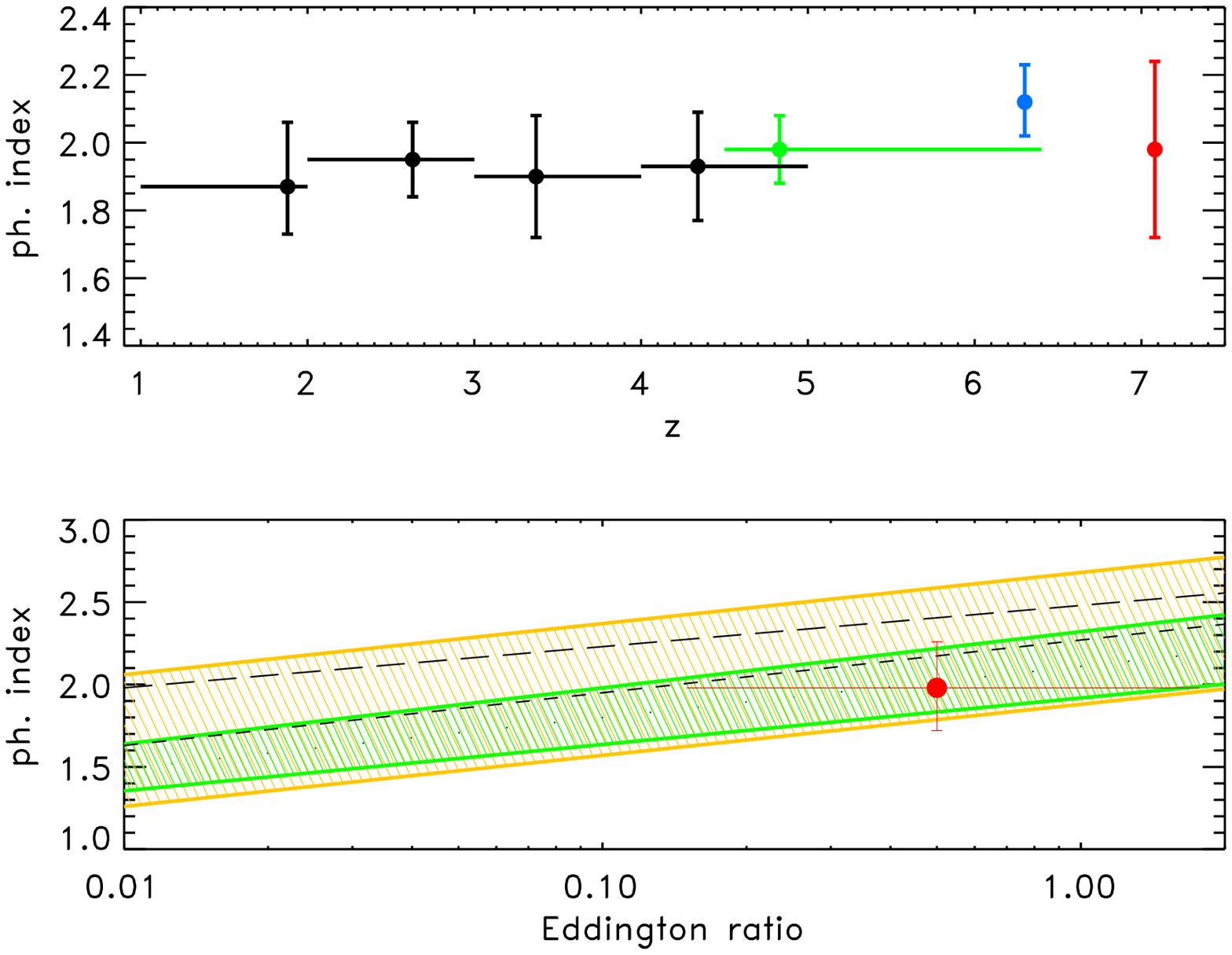}   
\caption{{\bf Upper panel:} photon index versus redshift. Black and green points are resul1.ts of stacking analysis 
from \cite{Just07} and \cite{Vignali05}; the blue point is the photon index of SDSS J1030+05 \citep{Farrah04};
the red point is our measure of the photon index of ULASJ1120+0641. 
In all the cases the assumed  model is a simple power law.  
{\bf Lower panel:} photon index versus Eddington ratio. The yellow area represents the correlation
together with its scatter found by \cite{Risaliti09} while the green shaded area is from \cite{Shemmer08}.
The short and long  dashed line are  from \cite{Brightman13} and \cite{Fanali13}, respectively.
Red point shows the present measure of the photon index of ULASJ1120+0641, with the Eddington ratio estimate by
\cite{Derosa13}. The error bar is calculated accounting for the zero point uncertainty (0.55 dex) of the relation used to compute the BH mass \citep{Derosa13}.}
\label{fig:gam}
\end{figure}
The typical AGN X-ray spectrum above 2 keV can be well described by a simple power law with a photon index 
of $\sim$ 2 with a 10-20\% dispersion either considering only data below 10 keV \citep[e.g.]{Tozzi06,Mainieri07,Young09,Corral11}
or including higher energies \citep[e.g]{Burlon11,Rivers13}.
As several previous studies \citep{Vignali05, Shemmer06, Just07}  did not find  any detectable evolution of the AGN 
photon index change with redshift up to z $\sim$ 5, we expect that  ULAS1120+0641 photon index would be consistent with the collective properties of lower redshift populations (upper panel of Fig.\ref{fig:gam}). Indeed the measure of 1.98$^{+0.26}_{-0.28}$ is well consistent with the typical 2.0 and a 10-20$\%$ scatter . 
The only other  direct measure of a z$>$6 AGN X-ray photon 
index is $\sim$ 2.1$\pm$0.1 \citep{Farrah04}, while the stacking analysis
of a sample of 24  z$>$4.5 quasars yields 1.98$\pm$0.2 \citep{Vignali05}.

A relation between the Eddington ratio and X-ray spectral slope has been found by several 
authors using different samples \citep{Shemmer06,Risaliti09,Jin13,Fanali13,Brightman13};
see the lower panel of Fig.\ref{fig:gam}.
A plausible physical interpretation of this correlation is that, at high accretion rates, 
the disk emission is stronger and  produces more soft photons leading to a more efficient 
Compton cooling of the corona and, therefore, to steeper X-ray spectra. 
ULAS1120+0641, with an  estimated Eddington ratio of 0.5 \citep{Derosa13},  according 
to \cite{Shemmer08}, \cite{Risaliti09} and \cite{Fanali13},
is expected to have an X-ray spectrum slope of  2.0-2.4,
which is within the 1$\sigma$ uncertainty of  the actual measure (lower panel of Fig.\ref{fig:gam}). 
\subsection{X-ray and UV} 
\label{sect:xuv} 
\begin{figure}
\includegraphics[width=\columnwidth]{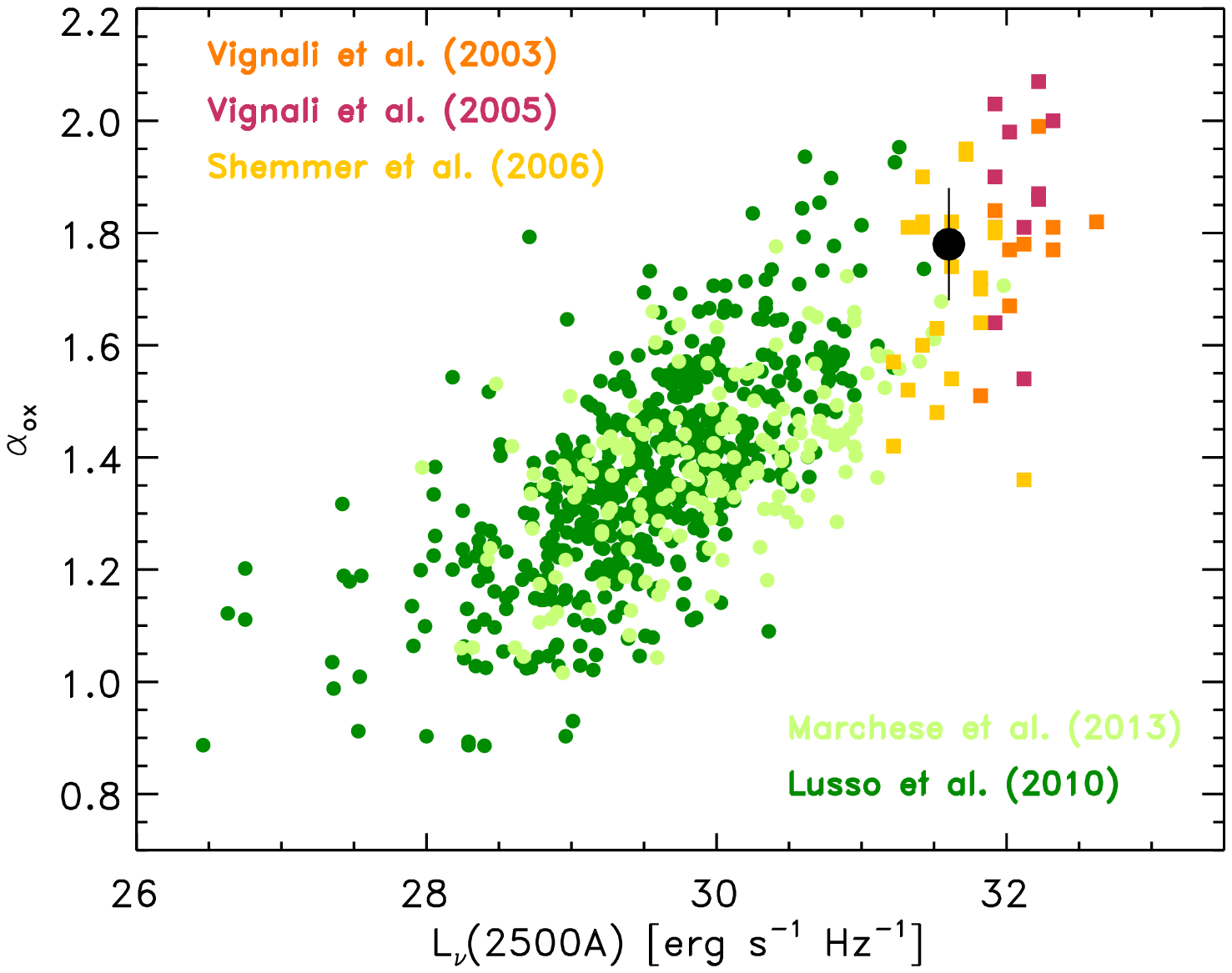}   
\caption{The $\alpha_{\rm OX}$ versus UV monochromatic luminosity of ULAS1120+0641(black point), compared with four different 
samples, both  X-ray \citep{Lusso10, Marchese12} and optically selected \citep{Vignali03,Vignali05,Shemmer06}}
\label{fig:aox}
\end{figure}

The ratio between X-ray and UV-optical emission is expected to be determined by the relative importance 
between disk and corona. Therefore it is an important piece of information to test energy generation models and to constrain 
the bolometric corrections. 
From a purely observational point of view, determining this ratio is mandatory to compare statistical 
properties of samples which have been selected at different wavelengths. 
The ratio between the UV and X-ray emission is usually parametrized by  
$\alpha_{\rm OX}$
\footnote{defined as $\alpha_{\rm OX}=-$log(f$_{2500\rm \AA}$/f$_{2\rm keV}$)/log($\nu_{2500\rm \AA}$/$\nu_{2\rm keV})$.} which is the slope of the power law that joins the energy distribution at 2500$\rm \AA$ and 2 keV.

For ULAS1120+06 we measured  a 2 keV rest-frame monochromatic luminosity of  7.8$^{+5.2}_{-3.1}$$\times$10$^{26}$ erg s$^{-1}$ Hz$^{-1}$,
while from the IR spectrum \citep{Mortlock11} we derived a 2500 $\AA$ rest-frame  luminosity of  3.4 $\times$10$^{31}$ erg s$^{-1}$ Hz$^{-1}$,
assuming a 10\% uncertainty.  This yields an $\alpha_{\rm OX}$=1.78 $^{+0.10}_{-0.10}$. 

\cite{Vignali03b}, \cite{Steffen06} and,  more recently, \cite{Lusso10} found that the apparent dependence of 
 $\alpha_{\rm OX}$ on redshift can be explained by 
a selection bias, and confirmed that the real  dependence is on the UV luminosity. 
Indeed all these authors found a significant correlation between   $\alpha_{\rm OX}$ and the monochromatic L$_{\nu}$(2500A)
in the sense that the X-ray  luminosity is relatively lower  for UV brighter AGN (Fig. \ref{fig:aox}). 
In agreement with these results, we find that ULAS1120+06 $\alpha_{\rm OX}$ is consistent with lower redshift (4$<$z$<$5) 
optically selected  bright QSO I \citep{Vignali03, Vignali05, Steffen06, Just07} which have similar UV luminosities.
Moreover, we find very good consistency with the extrapolation to  bright UV luminosities of the  L$_X-$L$_{UV}$ 
correlation as measured in the X-ray selected AGN COSMOS \citep{Lusso10} and XBS  \citep{Marchese12} samples (Fig. \ref{fig:aox}).

Accordingly with what is found for the $\alpha_{\rm OX}$ parameter, we note that the value of  the X-ray bolometric correction, k$_{\rm bol}$, defined
as the ratio between the X-ray and bolometric luminosities (361$^{+19}_{-15}$), is well consistent 
with the expectation for quasars at this level of X-ray loudness \citep{Lusso10,Marchese12}.
In particular, using the relation found by \cite{Marchese12}, log (k$_{\rm bol}$) = 1.05 - 1.52$\alpha_{\rm OX}$+1.29$\alpha_{\rm OX}^2$ we expect a bolometric correction of 387 with a 1 $\sigma$ scatter of 54 (corresponding to 14\%), which is in very good agreement with our measure.

\noindent

\section{Conclusion}
In this paper we report on the deep X-ray observation of the quasar ULAS1120+0641. 
This is the  second time that the  X-ray spectrum, and, in particular, the  direct measure of the  photon index of
an AGN at redshift z $>$6 can be measured.
We found that the X-rays contribute only for the 0.3$\%$ to the bolometric energy
output of this source. 
Both the X-ray spectrum and the X-ray-to-optical spectral index are consistent with the properties 
of bright quasar samples optically detected at lower redshift. 
This suggests that the physical mechanism, which is behind the broadband emission 
of AGNs in the local Universe is already active in ULAS1120+064, only 0.75 Gyr after the Big Bang.

As pointed out by \cite{Derosa13}, the systematic uncertainties in the mass determination
(0.3-0.55 dex) prevent from placing robust constraints on the masses of the SMBH seeds.
In this context the measure of the X-ray spectrum slope provides a useful extra piece of information
since we know that this slope is related to the Eddington ratio.
Our work shows that this is the case also for ULAS1120+064, for which the X-ray spectrum is consistent
with the Eddington ratio independently determined from the optical spectrum.

In the last few years, wide area optical/IR photometric surveys have been able to 
extend the AGN studies at very high redshift  well inside the re-ionisation era 
and when the coevolution with their host galaxies has been settling.
This has been possible thanks to the  detection of  very bright and rare quasars,
which represent the most striking sources of the AGN population,   
the bulk of which is much fainter.
In this context X-ray observations would be mandatory to probe the obscured accretion.
However, building an X-ray selected catalog of z$>$6 AGNs is beyond the capabilities of the 
current generation of X-ray missions, which are limited by a relatively small grasp 
(effective area $\times$ field of view).
The wide field imager (WFI),  which will be part of the scientific payload of the Athena+ mission \citep{Nandra13}, 
recently selected as  second large class mission (L2) in the Cosmic Vision program,
will improve the survey velocity by two orders of magnitude with respect to Chandra.
This will allow to detect a relevant number (60) of z$>$6 AGN employing the same exposure time of the 
Chandra deep field, four millions seconds, \citep{Aird13}.

At the present time, X-ray follow-up observations of optically/IR selected quasars are useful  not only to calculate the bolometric luminosity and  improve the accretion rates estimates, but also to  test possible evolution of the physical mechanism acting very close to the SMBH.
For high redshift AGNs this is very important as the measure of the mass, together with an estimate of the accretion rates 
for a statistically significant sample, can provide the constrains to distinguish between different model of SMBH formation
and evolution \citep{Volonteri10}.

\begin{acknowledgements}
The research leading to these results has received funding from the European Commission Seventh Framework Programme (FP7/2007-2013) under grant agreement n.267251 "Astronomy Fellowships in Italy" (AstroFIt).
\end{acknowledgements}
\bibliographystyle{aa}   
\bibliography{/Users/alberto/scripta/papers/tot09}   
\end{document}